\documentclass[twocolumn]{article}

\usepackage{graphicx}
\usepackage[doi=false,url=false,isbn=false,sorting=none]{biblatex}
\usepackage{microtype}
\usepackage[top=1in, bottom=1in, left=0.8in, right=0.8in]{geometry}
\usepackage{color}
\usepackage{colortbl}
\usepackage{caption}
\usepackage{supertabular}
\usepackage{booktabs}
\usepackage{longtable}
\usepackage{multirow}
\usepackage{etoolbox}
\usepackage{tikz}

\graphicspath{{./images/}}

\title{\bfseries\fontsize{22}{1}\selectfont BEEBS: Open Benchmarks for Energy Measurements on Embedded Platforms}
\author{
\parbox{0.33\linewidth}{\centering\fontsize{13}{1}\selectfont James Pallister\\[0.5em]\texttt{jp@cs.bris.ac.uk}}
\parbox{0.33\linewidth}{\centering\fontsize{13}{1}\selectfont Simon Hollis\\[0.5em]\texttt{simon@cs.bris.ac.uk}}
\parbox{0.33\linewidth}{\centering\fontsize{13}{1}\selectfont Jeremy Bennett\\[0.5em]\texttt{jeremy.bennett@embecosm.com}} \\[1.5em]
\parbox{0.66\linewidth}{\centering\fontsize{12}{1}\selectfont
Department of Computer Science, \\
University of Bristol, \\
Merchant Venturers Building, \\
Woodland Road, Bristol, BS8 1UB\\
United Kingdom}
\parbox{0.33\linewidth}{\centering\fontsize{13}{1}\selectfont
Embecosm \\
Palamos House \#104, \\
66/67 High Street, \\
Lymington, SO41 9AL, United Kingdom. \\
}
}
\date{}

\bibliography{library}

\begin{document}
\maketitle

\section{Abstract}
{
\bf This paper presents and justifies an open benchmark suite named BEEBS, targeted at evaluating the energy consumption of embedded processors.

We explore the possible sources of energy consumption, then select individual benchmarks from contemporary suites to cover these areas. Version one of BEEBS is presented here and contains 10 benchmarks that cover a wide range of typical embedded applications. The benchmark suite is portable across diverse architectures and is freely available.


The benchmark suite is extensively evaluated, and the properties of its constituent programs are analysed.
Using real hardware platforms we show case examples which illustrate the difference in power dissipation between three processor architectures and their related ISAs. We observe significant differences in the average instruction dissipation between the architectures of 4.4x, specifically 170$\mu$W/MHz~(ARM Cortex-M0), 65$\mu$W/MHz~(Adapteva Epiphany) and 88$\mu$W/MHz~(XMOS XS1-L1).

}
\section{Introduction}

Benchmarking is frequently used to gain an idea of how a system will perform during general use, when the specific environment cannot be reproduced at design-time. This gives designers feedback on how their system will perform and where performance is lacking. Typically, one benchmark cannot exercise all aspects of a target, leading to suites of benchmarks. Each benchmark tests a combination of areas of the hardware. This separation of benchmarks allows the designer to see which parts of the hardware perform the best.

The energy consumption of electronic devices is rapidly becoming a large factor in the design process. A portable embedded system will typically have severe power constraints placed upon it, if it is to have a long battery life. To recognize whether these constraints have been met, the power consumption of the device under a typical load must be tested. To build a full picture of a platform's energy consumption characteristics, a benchmark suite that hits possible combinations of an application's characteristics (such as memory accesses, integer and floating point operations, etc) is needed. This allows the energy consumption of various components of the system to be determined, ensuring that the system is fit for purpose.

\begin{table*}[th!]
\centering
	\begin{tabular}{l c c c c c c l}
	\toprule
	Name				 	& Source 	& B & M & I & FP 		& License & Category \\
	\midrule
	Blowfish				& MiBench 	& L & M & H & L & GPL	& Security	\\
	CRC32					& MiBench 	& M & L & H & L & GPL	& Network, telecomm	\\
	Cubic root solver		& MiBench 	& L & M & H & L & GPL	& Automotive	\\
	Dijkstra				& MiBench 	& M & L & H & L & GPL	& Network	\\
	FDCT					& WCET 		& H & H & L & H & $\textrm{None}^\dagger$	& Consumer	\\
	Float Matmult			& WCET 		& M & H & M & M & $\textrm{None}^\dagger$	& Automotive, consumer	\\
	Integer Matmult			& WCET	 	& M & M & H & L & $\textrm{None}^\dagger$	& Automotive	\\
	Rjindael				& MiBench 	& H & L & M & L & GPL	& Security	\\
	SHA						& MiBench 	& H & M & M & L & GPL	& Network, security	\\
	2D FIR					& DSPstone	& H & M & L & H & $\textrm{None}^\dagger$	& Automotive, consumer	\\
	\bottomrule
	\end{tabular}
\caption{Benchmarks selected, and the categories they fit in. Legend in Table~\ref{BenchmarkLegend}. \\[-0.em]
 {\null\hfill\scriptsize $\dagger$ Redistributed under the GPL.\hspace{2cm}\null}
 }
\label{Table:BenchmarkTable}
\end{table*}

There are few freely available benchmark suites for deeply embedded systems and none exist which are designed to allow energy consumption to be measured. Existing suites, such as MiBench~\cite{Guthaus2001}, MediaBench~\cite{Fritts2009}, LINPACK~\cite{Dongarra2003a} and Dhrystone~\cite{Weicker1988} are all targeted towards larger desktop-based applications, with significant compute power. This is due to their emphasis on measuring performance, as opposed to energy efficiency. Most assume a host operating system is present, which may not be true on an embedded system. Furthermore, when analysing energy consumption, having to account for the operating system’s effect on the result is non-trivial. These benchmarks --- while in theory are portable --- have significant difficulties running unmodified on embedded platforms. There are a variety of issues that cause these difficulties, such as lack of an OS, lack of a storage system, small memory size and run-time scalability. The issue of run-time scalability only occurs with a diverse range of platforms --- large differences in clock speed and microarchitecture may mean that without scaling down a benchmark it is infeasible to run it on less powerful platforms.

Of the existing suites, MiBench is the closest to our requirements in terms of variety of benchmarks and applicability but assumes there is a host operating system for the majority of the benchmarks. In particular it requires access to a filesystem which is usually unavailable on small embedded platforms. The benchmarks represent a broad range of embedded areas. Our benchmark suite keeps this cross-section of areas while selecting benchmarks which bring out a range of energy consumption characteristics.

The WCET benchmarks~\cite{Gustafsson2010} are also quite suitable, in that none of them require an operating system. However, many of these programs are small and not representative of computations that would typically be done on an embedded platform (e.g. searching for primes).

The DSPstone suite~\cite{Zivojnovic} is aimed at evaluating compilers for DSP-type platforms, therefore it fits into the criteria of no OS and small memory footprint. However the majority of these benchmarks are too small to be useful in a realistic benchmark set.

In this paper we create a new set of benchmarks --- the Bristol Energy Efficiency Benchmark Suite (BEEBS)~\cite{JamesPallisterSimonHollis2013} --- chosen from popular benchmark suites, and their use justified for benchmarking energy consumption. The benchmark suite is designed to expose the processor and memory's performance, with other factors such as I/O and peripherals excluded for portability. The selection was designed such that the benchmarks would be portable, to expose the changing in energy consumption when exercising the platform in different ways, such as with memory verses arithmetic intensive computation. The benchmarks are intended to be run on the ‘bare metal’ with no host operating system.

We consider four orthogonal aspects that the benchmark suite must cover, allowing the range of benchmarks to expose all of the behaviour of the platform.

\begin{itemize}
	\setlength{\itemsep}{-0.em}
	\vspace{-2mm}
	\item Integer operations. Operations which use the integer ALU will have similar energy consumptions.
	\item Floating point operations. These operations may use different pipelines or functional units to the integer operations, so may consume a different amount of energy.
	\item Memory access intensity. An access to memory is known to take a significantly different amount of energy to other operations~\cite{Tiwari1994a}.
	\item Branching frequency. Branching frequently will stress parts of the processor, such as an instruction prefetch phase. This is similar to memory access intensity, but as the code and data are often held in different areas and types of memory this should be considered separately.
\end{itemize}

Using benchmarks that hit combinations of these, interesting observations about the energy consumption of the device can be made.

The benchmark suite has been extensively tested on three different processors, with the rest of the paper detailing the results, shown in the top half of Table~\ref{tab:platform_list}. The suite has been confirmed to run successfully on a further three platforms (shown in the bottom half of Table~\ref{tab:platform_list}). Targeting multiple platforms ensures that more general conclusions can be drawn about the nature of the energy consumption.

This paper discusses previous benchmark suites, justifying the need for a benchmark suite targeted at exposing energy consumption characteristics. Then a set of benchmarks chosen from subsets of these pre-existing suites is presented, with justifications listed for the benchmarks and the modifications made to them. An analysis of the new BEEBS suite is given, with instruction distributions and examples of how the benchmarks can be used to expose energy consumption characteristics.

\begin{table}[t]
\centering
	\begin{tabular}{c l}
		\toprule
		Key & Description \\
		\midrule
		L	&	Low \\
		M	&	Medium \\
		H	&	High \\
		\midrule
		B	&	Branching \\
		M	&	Memory intensity \\
		I	&	Integer pipeline intensity \\
		FP	&	FPU pipeline intensity \\
		\bottomrule
	\end{tabular}
	\caption{Legend for the benchmark table}
	\label{BenchmarkLegend}
\end{table}
\begin{table}
	\centering
	\begin{tabular}{l l}
		\toprule
		Vendor  & Processor \\
		\midrule
		ARM		  & Cortex-M0 \\
		Adapteva  & Epiphany \\
		XMOS	  & L1 \\
		\midrule
		ARM		  & Cortex-M3 \\
		ARM 	  & Cortex-A8 \\
		Microchip & PIC32MX (MIPS) \\
		\bottomrule
	\end{tabular}
	\caption{Platforms considered for the benchmark suite. The top half of the table is analysed in depth, where as the suite is verified to compile and run on the lower half.}
	\label{tab:platform_list}
\end{table}

\section{Previous Work}

Of the many existing benchmark suites, few target embedded systems. Most target either desktop machines (e.g Dhrystone) or HPC (e.g. PARSEC). Few also explicitly target multithreaded systems, and none explicitly aim for energy as the target metric.

MiBench established a well known set of benchmarks with well characterised behaviour. This suite consisted of 37 different benchmarks split across six different categories, chosen to be representative of which applications would be run on both desktop and embedded platforms. Each benchmark is justified, with instruction traces analysed on a model of the StrongARM architecture. This gave a good overview of the proportions of each type of instructions that the benchmarks executed. The drawback of this was that the instruction traces were only gathered for one platform --- each benchmark could have a radically different instruction distribution for alternative platforms, leading to a different performance characteristics exposed.

MiBench was used as the main benchmark suite for MILEPOST GCC~\cite{Fursin2011}. This study applied machine learning to predict which optimizations would benefit a program without needing to perform expensive iterative compilation techniques. In this study they emphasised how the performance achieved can be very dependent on the structure of the benchmarks. This highlights the need to have a wide range of benchmarks which each hit different combinations of the types of computation they could perform.

ParMiBench, a variant of MiBench was created to address the lack of multithreadedness in the original suite~\cite{Iqbal2010a}. It attempts to parallelise some of the benchmarks, allowing them to be used to benchmark multicore systems. This has an advantage over other parallel benchmark suites in that it also targets the embedded space. Very few other benchmark suites (such as LINPACK, PARSEC and SPLASH-2~\cite{Bienia2008}) target multithreadedness at this level --- most are aimed at large clusters and HPC applications.

DSPstone is a benchmark suite for Digital Signal Processors (DSPs) and was originally designed to evaluated compiler effectiveness at compiling for DSPs. This suite contains a large number of non-integer tests, with most tests replicated in fixed point and floating point form. As this set is aimed at DSPs rather than general purpose processors no benchmarks were chosen from DSPstone.

A set of benchmarks is maintained by the worst case execution time (WCET) initiative. These benchmarks are appropriate because they are self contained and written completely in C. Each benchmark is less comprehensive than its equivalent from the MiBench set, but focusses on one particular application that may be specifically what a low end processor will perform. Some of these applications fit well with typical embedded applications.

In addition to the previous benchmark suites, several other suites were evaluated.
We also evaluated several unsuitable suites:
\begin{itemize}
	\setlength{\itemsep}{-0.25em}
	\item MediaBench
	\item OpenBench\cite{Rebe2012}
	\item SPEC2006\cite{Henning2006}
	\item LINPACK
	\item Livermore Fortran Kernels
\end{itemize}

All of these benchmark suites were found to be unsuitable for the aim of characterising energy consumption on embedded platforms due to their reliance on the operating system and features provided by it.

A specific suite to target energy consumption is useful because of the differing energy costs of each instruction in a processor's instruction set. Many previous studies \cite{Blume2006,Lee2001,Steinke2001,Tiwari1996} have attached an energy cost to each instruction and find that different instructions can have significantly different energies even if they take a similar amount of time.

Brooks et al. created the \textit{Wattch} toolkit~\cite{Brooks2000} which provides architectural models and instruction level models to allow design-space exploration of the power consumption of processors, as well as evaluating software's energy consumption. BEEBS provides the missing component, a benchmark suite designed for energy exploration that allows these kind of explorations to be done consistently and systematically.

Energy modelling has also been used to optimise a program's execution, through selecting compiler optimisations~\cite{Patyk2009}, instruction scheduling~\cite{Parikh} and automatically inserting idle instructions~\cite{Seth2001}.

Optimisation can also be achieved at the microarchitecture level, for example, by choosing an instruction encoding to minimise the number of bit flips~\cite{Woo2001}. Other methods of reducing energy in this way include encoding bus traffic~\cite{Stan1995}, adaptive scheduling of DRAM accesses~\cite{Hur2008} and exposing energy efficient version of instructions in an ISA~\cite{Asanovic2000}.

\section{Platforms}

We intend BEEBS to be applicable to a wide range of hardware platforms. For our evaluation, a range of platforms has been chosen, covering different types of architectures. The processors are mainly small embedded systems which are designed for low power usage. As a consequence, some of the platforms are very memory limited, restricting the types of applications that can be run on them.

A set of platforms is needed to complement the benchmark suite due to the varying capabilities of each platform. For example, a benchmark will behave very differently on platforms which have a cache, compared to platforms which do not. As such, we have chosen platforms with different pipeline depths, numbers of registers and types of memory. A comparison of the platforms can be seen in Table~\ref{tab:platforms}.

The number of registers has a large effect on the energy consumption due to the high cost of memory accesses --- if a variable can be stored in a register there will be fewer memory accesses and overall less energy consumed. For similar reasons the type of memory the code is executing can have a large impact on energy --- flash and SRAM both consume different amounts of energy.

The XMOS platform is an unusual platform, in that it is an event driven multicore platform, with eight hardware threads. Of these threads, up to four can run full speed~\cite{Kerrison2013}. The Epiphany platform is superscalar having one integer pipeline and another integer/floating point pipeline. The Epiphany processor used has 16 cores, connected by a network-on-chip~\cite{Adapteva2013}. The ARM Cortex-M0 is a simple single-core processor.

All three platforms also have diverse instruction sets with different features. This diversity makes this selection of platforms ideal for testing the benchmarks.

\begin{table}
	\begin{tabular}{l l p{1cm} l p{1.8cm}}
		\toprule
		Platform & Registers & Pipeline depth & FPU & Execution memory\\
		\midrule
		Cortex-M0 & 16 & 3 & No & Flash \\
		XMOS L1 & 12 & 4 & No & RAM \\
		Epiphany & 64 & 8 & Yes & RAM \\
		\bottomrule
	\end{tabular}
	\caption{Features of the platforms experimented on.}
	\label{tab:platforms}
\end{table}

\section{The BEEBS Benchmarks}

A set of benchmarks to tests all aspects of the target platforms is presented in this section. The benchmarks were selected by defining a coverage matrix which included all the individual benchmarks from following suites:
\begin{itemize}
	\setlength{\itemsep}{-0.35em}
	\item MiBench
	\item DSPstone
	\item WCET
	\item Livermore Fortan Kernels
	\item Dhrystone
	\item MediaBench
\end{itemize}

The matrix (listed in full in Appendix~\ref{appendix:benchmarks}) also broadly evaluated other benchmark suites for their suitability. Two sets of parameters are evaluated in this table --- type of operations performed by the benchmark and suitability for inclusion in the final suite. The suitability for inclusion evaluates whether the benchmark should be included, based on what the benchmark does, whether it will work on the target platforms and the effort required to port it.

The type of operations was derived from examining the source of each benchmark and roughly categorising it as to the types of operations it performs. This allows benchmarks with similar properties to be excluded before a lengthy examination.

Benchmarks with a high suitability and a minimal set covering suitably different types of operations were selected to be included in the final suite (shown in Table~\ref{Table:BenchmarkTable}). The types of operations are listed were calculated from a combination of inspecting the source code and from the instruction traces generated. This is shown in the table under the following columns:
\begin{itemize}
	\setlength{\itemsep}{-0.35em}
	\vspace{-2mm}
	\item \textbf{B}ranching.
	\item \textbf{M}emory.
	\item \textbf{I}nteger.
	\item \textbf{F}loating \textbf{P}oint.
\end{itemize}

In the final suite, a large number of benchmarks are derived from MiBench. MiBench has 37 well defined benchmarks, however a large proportion of these are targeted at much higher end platforms than chosen. This lead to a small subset of the MiBench benchmarks being selected. Several benchmarks were sourced from the WCET set. These tested small applications which could conceivably be ran by the platforms discussed earlier. One benchmark is taken from the DSPstone suite, to cover this application area and type of computation.

The other applications considered were all found to be too time consuming to port to a small embedded system, or unnecessary for inclusion because other benchmarks performed a similar set of operations.

\section{Benchmark Descriptions}

This section talks about each benchmark, giving a short description of the benchmark, modifications made, and why it is included.

\subsection*{Categories}

MiBench divided the embedded processor applications into six categories (see Table~\ref{tab:categories}): automotive, network, consumer, security, telecomms and office. The benchmarks selected broadly fit into these categories, however consumer and office in particular require the higher end embedded processors. This is due to the benchmarks running `off the shelf' programs such as ghostscript and rsynth.

Similarly we divide the chosen benchmarks into the same categories, since they are appropriately descriptive. However, some of the benchmarks are broad enough that the fit into several categories. A more accurate classification of the groups the benchmarks fit into is shown in the table of benchmarks (Table~\ref{Table:BenchmarkTable}).

\begin{table}
	\centering
	\begin{tabular}{l p{0.55\linewidth}}
		\toprule
		Category &	Description \\
		\midrule
		Automotive 	& This category demonstrates the mathematical ability of the processor. \\
		Consumer	& Embedded processors are frequently used in consumer applications, performing tasks such as audio and video decoding. \\
		Network		& Processors in routers frequently perform the operations in this category. This involves handling packets and routing graphs. \\
		Telecomm	& Applications that include radio frequency analysis and encoding. \\
		Security	& Encryption algorithms, hashing and signing applications are placed in this category. \\
		\bottomrule
	\end{tabular}
	\caption{Categories used to indicate the application area of the benchmarks.}
	\label{tab:categories}
\end{table}

\subsubsection*{Blowfish}
Blowfish is an encryption algorithm commonly used in cryptography. This benchmark was taken from MiBench but modified to both encrypt and decrypt small blocks of data, as if the data was being streamed into the processor. The stream is generated pseudo-randomly to avoid platform dependencies on input and output. Encryption typically involves many integer operations with fewer, predictable branches.

\subsubsection*{Rijndael}
Rijndael is the algorithm for the Advanced Encryption Standard. It is commonly used in many security applications, and has a similar structure to blowfish. It also has similar execution characteristics except for more frequent branching.

\subsubsection*{SHA}
Secure Hashing Algorithm (SHA) is a hashing algorithm commonly used for fingerprinting and verification of data streams. It is useful for stressing integer pipelines, and has low memory requirements. The benchmark hashes a stream of pseudo randomly generated data.

\subsubsection*{CRC32}
Similar to SHA, CRC32 is used for verification of data streams, notably ethernet frames. It differs from SHA in that it can be implemented with very few instructions as it consists mainly of shifts and XORs. As it consists of few instructions in a tight loop, this benchmark should exercise processors with superscalar execution or branch prediction. The benchmark performs the CRC on a stream of pseudo randomly generated data.

\subsubsection*{Integer Matrix Multiplication}
Integer matrix multiplication is used very frequently in many applications, and so is a useful benchmark to have. It consists of a tight inner loop with many array accesses, making it useful for stressing the memory and integer pipeline at the same time. This should also expose data caching effects of the platform.

\subsubsection*{Float Matrix Multiplication}
Floating point matrix multiplication is also used frequently. This benchmark is a modified version of the integer matrix multiplication benchmark, with floating point numbers in place of integer --- all other code is identical. This should allow a good metric of relative performance between the integer and floating point pipeline to be produced.


\subsubsection*{Dijkstra}
This benchmark implements the Dijkstra shortest-algorithm path. This benchmark performs lots of non-linear accesses to memory, and branches unpredictably. This makes it good for stressing caches and branch units that the processor may have. This algorithm is commonly used by routers to calculate the shortest path to another router. This benchmark was modified from the MiBench version to have the adjacency matrix embedded in the source code, rather than loaded from the filesystem.


\subsubsection*{Cubic root solver}
This benchmark performs a large amount of trigonometry to solve various cubic equations. This tests the floating point pipeline with very little memory required. This is a portion of the `basicmath' benchmark in MiBench, cut down to fit on smaller processors.

\subsubsection*{2D FIR}
FIR filters are frequently using in image transformations. In the embedded space this could be the type of operations done by digital cameras. This benchmark is similar to the matrix multiplications but with potentially more memory accesses and spatially different arithmetic.


\subsubsection*{FDCT}
The Finite Discrete Cosine Transform (FDCT) benchmark was included as it is a core algorithm behind many video decoders used in consumer products. This benchmark represents real-world usage of the systems as well as testing the floating point pipeline and caches.

\section{Benchmark Analysis}

\begin{figure}[t!]
	\centering
	\includegraphics[width=0.9115\linewidth,clip,trim=1cm 0.5cm 0cm 1.2cm]{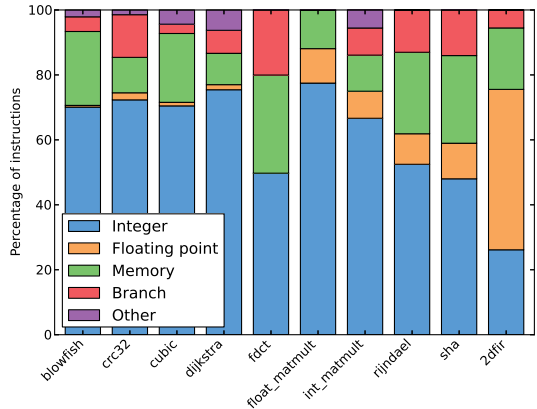}
	\caption{BEEBS Instruction distribution for the Epiphany platform.}
	\label{Fig:InstructionDistributionEpiphany}
\end{figure}

\begin{figure}[t!]
	\centering
	\includegraphics[width=0.9115\linewidth,clip,trim=1cm 0.5cm 0cm 1.2cm]{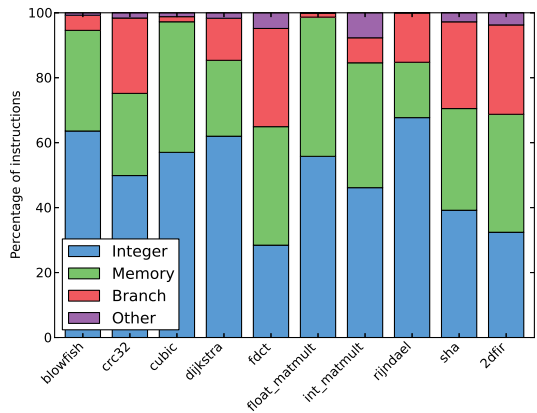}
	\caption{BEEBS Instruction distribution for the XMOS platform.}
	\label{Fig:InstructionDistributionXMOS}
\end{figure}

\begin{figure}[t!]
	\centering
	\includegraphics[width=0.9115\linewidth,clip,trim=1cm 0.5cm 0cm 1.2cm]{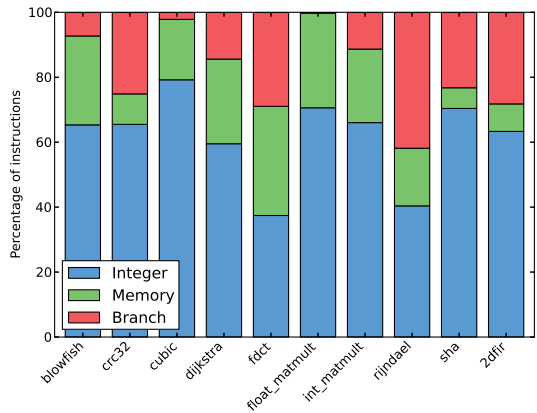}
	\caption{BEEBS Instruction distribution for the ARM Cortex-M0 platform.}
	\label{Fig:InstructionDistributionARM}
\end{figure}

This section provides a concrete analysis of all the chosen benchmarks by collecting their instruction traces across three of the platforms. From these graphs, the instructions can be categorised to demonstrate that each benchmark performed a different distribution of operations. Figures~\ref{Fig:InstructionDistributionEpiphany}, \ref{Fig:InstructionDistributionXMOS} and \ref{Fig:InstructionDistributionARM} show the instruction distributions for the Epiphany, XMOS and ARM Cortex-M0 (Thumb instruction set) platforms respectively. The `Other' category of instructions contains miscellaneous control instructions that do not fit into other categories (for example, interrupt control on the Epiphany platform).

Overall these results show that the benchmarks give a good spread of different distributions of instruction types.

Integer operations are the most common type of instruction in almost every benchmark. Across the platforms, the distributions are similar, with small variations due to the underlying instruction set. For example, there are a larger percentage of \texttt{mov}-type instructions in the Epiphany results because there are several predicated \texttt{mov} instructions (\texttt{moveq}, \texttt{movlt}, etc). This reduces the need for conditional branches, so this category decreases in proportion.

\begin{table}
	\centering
	\begin{tabular}{>{\bf}c c r @{--} l  r @{--} l  r @{--} l}
		\toprule
		Type & Platforms (\%) & \multicolumn{6}{c}{Benchmarks (\%)} \\
		     &           & \multicolumn{2}{c}{Epiphany} & \multicolumn{2}{c}{XMOS} & \multicolumn{2}{c}{ARM} \\
		\midrule
		I    & 30        & 26&77   & 28&68   & 37&79  \\
		FP   & --         & 0&49    &   &     &   &    \\
		M    & 30        & 10&30   & 17&43   &  6&34  \\
		B    & 29        & 1&20    &  1&30   &  1&42  \\
		\bottomrule
	\end{tabular}
	\caption{Variation in instruction distributions between the platforms and between the benchmarks.}
	\label{tab:inst_variation}
\end{table}

Epiphany is also the only platform in the subset chosen which has hardware support for floating point. For the other platforms, software emulation is used. On the XMOS platform this manifests in extra branch and memory instructions, whereas for the ARM platform the proportion of integer operations rises. These differences are due to different emulation strategies used.

The ARM traces follow the same general trend as the traces for XMOS and Epiphany, however with overall less memory operations. This is due to the ARM processor having support for the \texttt{ldm} and \texttt{stm} instruction allowing multiple accesses to memory in a single instruction. These instructions are used extensively in function prologues and epilogues to save and restore registers.

The integer instruction category is the largest group in almost every case, for all platforms and benchmarks. This comes from the integer category covering the largest number of types of instructions, as it groups arithmetic, register copying and bit-wise operations.

These benchmarks show a range of different quantities of each instruction, with similarities across platforms. This makes the set of benchmarks ideal for use in energy profiling of a system.

We see that for all platforms a given benchmark produces a similar instruction profile (within 30\% between all platforms). This is shown in Table~\ref{tab:inst_variation}, where the platforms column shows the maximum variation between each platform for each instruction category. The benchmark columns show the ranges of instruction proportions across the benchmarks on that platform. Between benchmarks there is significant variation, therefore the suite explores a wide range of input configurations in a consistent way between architectures.

\section{Case Study}
\label{sec:case_study}
\begin{table}
	\centering
	\begin{tabular}{l c c c}
		\toprule
						& \multicolumn{3}{c}{Power (mW)}\\
		Category 		& Epiphany 	& XMOS 	& ARM  	\\
		\midrule
		Integer         &   28     & 33    & 8.4  \\
		Floating Point  &   31     & --    & --  	\\
		Memory          &   20     & 35    & 9.3  \\
		Branching       &   40     & 35    & 6.8  \\
		Other           &   14     & --    & --     \\
		\midrule
		Average			&	26     & 35   & 8.3  \\
		\midrule
		Average/MHz		&   65$\mu$W & 88$\mu$W & 170$\mu$W   \\
		\bottomrule
	\end{tabular}
	\caption{Power dissipation for each instruction category calculated by linear regression.}
	\label{Table:RegressionResults}
\end{table}

\begin{figure}
	\centering
	\includegraphics{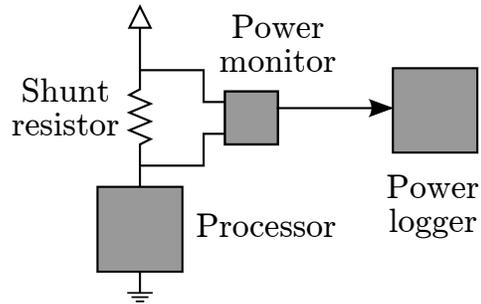}
	\caption{Hardware setup to measure the power of the processor under test.}
	\label{fig:hardware}
\end{figure}

The use of the benchmark suite is demonstrated through collecting power measurements for each benchmark on each of the platforms. Linear regression is then used to assign an average power dissipation to each class of instructions by considering the average power and instruction distribution per benchmark.

The power of each platform was measured by instrumenting hardware as in Figure~\ref{fig:hardware}. This set-up allowed real measurements to be taken, rather than using an abstract power model for the processor.

The average power dissipation of each benchmark was measured on the three hardware platforms. Linear regression is applied, with the categorized instruction counts gathered from the traces. This allows each category of instructions to be assigned an average power dissipation. The results of this analysis are presented in Table~\ref{Table:RegressionResults}. These are scaled results, representing the cost of a single instruction per core/hardware thread (Scaled down by 16 for Epiphany and by 4 for XMOS).

Overall, the main difference in power dissipations is due to differing clock rates --- XMOS and Epiphany run at 400MHz and ARM at 48MHz.

From these results several conclusions can be drawn. For the ARM Cortex-M0, a memory access is more costly than an arithmetic instruction, as is expected. The branch power dissipation, disagrees with other results taken. The power measured when executing a \texttt{while(1);} loop was found to be 11mW. This figure is higher than a memory access, due to the instruction being loaded from flash as opposed to RAM. The discrepancy is due to conditional branches having a lower power when the branch is not taken (further results indicate that when a conditional branch is not taken, the power dissipation is roughly 4mW).

The XMOS results show memory operations are slightly more costly than arithmetic. The identical cost for branching and memory access is due to the structure of the processor's pipeline: the final stage is a memory access which either does an instruction fetch or a memory operation.

The results for the Epiphany exhibit the most variability, with a branch instruction requiring almost twice the power of a memory access. We believe this is due to the longer pipeline having to be flushed, then new instructions fetched. A floating point operation also takes more power than an integer instruction --- this is attributed to the larger complexity of an FPU.

\section{Conclusion}

This paper presented BEEBS, a benchmark suite of 10 programs that has been carefully designed to expose the energy consumption characteristics of the target platform. The benchmarks were chosen after evaluating an extensive list of embedded programs for their characteristics and suitability. This included modifying existing benchmarks to be more suitable for a bare-metal benchmark suite for testing energy. The benchmarks are available online~\cite{JamesPallisterSimonHollis2013}.

Each of the benchmarks in the suite was analysed for its instruction distribution, verifying that the benchmark suite sufficiently covered the a range of distributions. This was repeated across three platforms with very different features, showing that the suite is consistently good even for different instruction sets. This is important when considering energy consumption, as each type of instruction can consume very different amounts of energy.

An example of how the benchmark suite could be used was given in Section~\ref{sec:case_study}. This case study took physical measurements of three platforms, ARM Cortex-M0, XMOS XS1-L1 and Adapteva Epiphany. Then an average power for each instruction was derived, by performing linear regression on the power figures and the instruction distributions. We find that different categories of instruction have different power consumptions, as expected. The power dissipations differ per platform in ways which can be explained. One example of this is the memory and branching consuming similar powers on the XMOS platform, due to the nature of the processor's pipeline. On the Epiphany platform floating point was slightly more power hungry and integer calculations, due to the extra circuitry in FPUs. The fact that these features can be highlighted by the suite shows that the benchmark suite is fit for purpose when evaluating different processors.

\subsection{Future Work}

The benchmark suite targeted the processor core of the embedded platforms, not exercising peripherals or I/O. In future this benchmark suite could be extended to allow these items to be tested, but it remains to be seen how this can be done in a portable way.

\printbibliography

\onecolumn
\appendix
\section{Benchmark Evaluation Table}
\label{appendix:benchmarks}

This appendix gives a comprehensive list of all the benchmarks evaluated to choose the final set. Each benchmark was examined and its rough characteristics estimated. Each benchmark was also evaluated for several other properties --- embedded applicability, memory footprint, and the modifications required to make it run on an embedded system. These three properties were combined in a rule-based manner, producing a `suitability' for inclusion in the suite. This allowed us to immediately exclude benchmarks with a very low suitability.

The characteristics of the benchmarks estimated the amount of computation in the following areas:
\begin{itemize}
	\item \textbf{I}nteger, \textbf{F}loating \textbf{P}oint, or neither. This was estimated from the ratio of floating point operations to integer operations.
	\item \textbf{B}ranching. The benchmark was deemed to be branch-intensive in there was a high ratio of control structures compared to other computation. For example, if more then 20\% of the code is control structures the benchmark was marked as branch-intensive.
	\item \textbf{M}emory. The benchmark was said to be memory intensive if there were frequent accesses to large arrays or other data structures.
\end{itemize}

The other columns in the table are:
\begin{itemize}
	\item \textbf{Embedded Applicability}. This is the likelihood that the functionality of the benchmark would be used in a real embedded system. For example, checksumming is frequently done in embedded systems, so this would receive a `High' embedded applicability.
	\item \textbf{Fit in memory}. This column specifies whether the benchmark would fit into a small amount of memory. Some benchmarks receive a `possibly' result for this, where it may be possible to reduce the size of the dataset the program uses.
	\item \textbf{Modifications for bare metal}. This field indicates the amount of modification necessary to make the benchmark run without operating system support. For example, if the benchmark does not make extesnive use of the operating system, and simply loads a dataset, the modifications to make this run bare metal are `minor'. However if the benchmark needs graphical display support or other complex features, the modifications necessary are `major'.
\end{itemize}

\vspace{5mm}

\begin{center}

\newcommand{\condcol}[1] {
	\hspace{-7mm}
	\ifstrequal{#1}{Very High}{\tikzstyle{hstyle}=[green!40!black!100]}{}
	\ifstrequal{#1}{High}{\tikzstyle{hstyle}=[green]}{}
	\ifstrequal{#1}{Medium}{\tikzstyle{hstyle}=[yellow]}{}
	\ifstrequal{#1}{Low}{\tikzstyle{hstyle}=[orange]}{}
	\ifstrequal{#1}{Very Low}{\tikzstyle{hstyle}=[red]}{}
	\tikz[overlay]\draw [fill, hstyle, draw=black,line width=0.2mm] (0,0) -- (0.25,0) -- (0.25,0.25) -- (0,0.25) -- (0, 0);
	\hspace{2mm}
	#1
	}
\definecolor{lightlightgray}{gray}{0.9}

\tablehead{%
    \toprule
    \bf \multirow{2}{*}{Benchmark} & \multicolumn{3}{c|}{\bf Characteristics} &\bf \multirow{2}{2.2cm}{Embedded Applicability} &\bf \multirow{2}{1.2cm}{Fit in memory} &\bf \multirow{2}{2.5cm}{Modifications for bare metal} &\bf \multirow{2}{*}{Suitability} \\
              &\bf  FP/I &\bf B &\bf M         &\bf                &               &            &             \\
    \midrule}
\tabletail{%
    \bottomrule}

\begin{supertabular}{l | c c c | l l l l}
\bf DSPstone  &&&&&\\
\midrule
Real updates       & FP &--&--& Medium   & Yes\footnotemark[1]  & None       & \condcol{High}\\
Matrix products    & FP &--&--& High     & Yes                  & None       & \condcol{Very High}\\
Complex product    & FP &--&--& Medium   & Yes\footnotemark[1]  & None       & \condcol{High}\\
LMS filter         & FP &--&--& Low      & Yes\footnotemark[1]  & None       & \condcol{Medium}\\
\rowcolor{lightlightgray}
2D FIR filter      & FP &--&Y& Medium   & Yes                  & None       & \condcol{High}\\
Complex updates    & FP &--&--& Medium   & Yes\footnotemark[1]  & None       & \condcol{High}\\
Convolution        & FP &--&--& Medium   & Yes\footnotemark[1]  & None       & \condcol{High}\\
IIR biquad filter  & FP &--&Y& Low      & Yes\footnotemark[1]  & None       & \condcol{Medium}\\
FIR filter         & FP &--&Y& Low      & Yes\footnotemark[1]  & None       & \condcol{Medium}\\
\midrule
&&&&&&&\\
\bf MiBench  &&&&&&&\\
\midrule
\rowcolor{lightlightgray}
basicmath          & FP &--&--& Medium   & Possibly   & None       & \condcol{Medium} \\
bitcount           & I  &Y &--& Medium   & Yes        & None       & \condcol{High} \\
qsort              & I  &Y &--& Medium   & Yes        & None       & \condcol{High} \\
susan (edges)      & FP &--&Y & Low      & Possibly   & None       & \condcol{Low} \\
susan (corners)    & FP &--&Y & Low      & Possibly   & None       & \condcol{Low} \\
susan (smoothing)  & FP &--&Y & Low      & Possibly   & None       & \condcol{Low} \\
jpeg               & -- &Y &Y & Medium   & No         & None       & \condcol{Medium} \\
lame               & FP &Y &Y & Low      & No         & None       & \condcol{Low} \\
mad                & -- &Y &Y & Low      & No         & Major      & \condcol{Very Low} \\
tiff2bw            & -- &Y &Y & Medium   & Possibly   & Major      & \condcol{Low} \\
tiff2rgba          & -- &Y &Y & Medium   & Possibly   & Major      & \condcol{Low} \\
tiffdither         & -- &Y &Y & Medium   & Possibly   & Major      & \condcol{Low} \\
tiffmedian         & -- &Y &Y & Medium   & Possibly   & Major      & \condcol{Low} \\
typeset            & -- &Y &--& Low      & No         & Major      & \condcol{Very Low} \\
ghostscript        & -- &Y &--& Low      & No         & Major      & \condcol{Very Low} \\
ispell             & -- &--&--& Low      & No         & Major      & \condcol{Very Low} \\
rsynth             & FP &--&--& Low      & No         & Major      & \condcol{Very Low} \\
sphinx             & -- &Y &--& Low      & No         & Major      & \condcol{Very Low} \\
stringsearch       & -- &Y &Y & Medium   & Yes        & None       & \condcol{High} \\
\rowcolor{lightlightgray}
dijkstra           & -- &Y &--& High     & Yes        & Minor      & \condcol{High} \\
patricia           & -- &Y &Y & High     & Yes        & Minor      & \condcol{High} \\
\rowcolor{lightlightgray}
blowfish enc       & I  &--&--& High     & Yes        & Minor      & \condcol{High} \\
\rowcolor{lightlightgray}
blowfish dec       & I  &--&--& High     & Yes        & Minor      & \condcol{High} \\
pgp sign           & I  &--&--& Medium   & Yes        & Minor      & \condcol{Medium} \\
pgp verify         & I  &--&--& Medium   & Yes        & Minor      & \condcol{Medium} \\
\rowcolor{lightlightgray}
rijndael enc       & I  & Y&Y & High     & Yes        & Minor      & \condcol{High} \\
\rowcolor{lightlightgray}
rijndael dec       & I  &Y& Y & High     & Yes        & Minor      & \condcol{High} \\
\rowcolor{lightlightgray}
sha                & I  &--&Y & High     & Yes        & Minor      & \condcol{High} \\
\rowcolor{lightlightgray}
CRC32              & I  &--&Y & High     & Yes        & Minor      & \condcol{High} \\
FFT                & FP &Y &Y & Medium   & Yes        & None       & \condcol{High} \\
IFFT               & FP &Y &Y & Medium   & Yes        & None       & \condcol{High} \\
ADPCM enc          & -- &Y &--& High     & Yes        & None       & \condcol{Very High} \\
ADPCM dec          & -- &Y &--& High     & Yes        & None       & \condcol{Very High} \\
GSM enc            & I  &--&--& High     & Yes        & None       & \condcol{Very High} \\
GSM dec            & I  &--&--& High     & Yes        & None       & \condcol{Very High} \\
\midrule
&&&&&&&\\
\bf WCET Benchmarks &&&&\\
\midrule
adpcom             & I  &--&--& High     & Yes        & None       & \condcol{Very High} \\
bs                 & -- &Y &--& Medium   & Yes        & None       & \condcol{High} \\
bsort100           & -- &Y &--& Medium   & Yes        & None       & \condcol{High} \\
cnt                & -- &--&Y & Low      & Yes        & None       & \condcol{Medium} \\
compress           & I  &--&--& Medium   & Yes        & None       & \condcol{High} \\
cover              & -- &Y &--& Low      & Yes        & None       & \condcol{Medium} \\
crc                & I  &Y &--& Medium   & Yes        & None       & \condcol{High} \\
duff               & -- &Y &--& Low      & Yes        & None       & \condcol{Medium} \\
edn                & I  &--&Y & Low      & Yes        & None       & \condcol{Medium} \\
expint             & I  &--&--& Medium   & Yes        & None       & \condcol{High} \\
fac                & -- &Y &--& Low      & Yes        & None       & \condcol{Medium} \\
\rowcolor{lightlightgray}
fdct               & I  &--&--& High     & Yes        & None       & \condcol{Very High} \\
fft1               & I  &--&--& High     & Yes        & None       & \condcol{Very High} \\
fibcall            & I  &Y &--& Low      & Yes        & None       & \condcol{Medium} \\
fir                & -- &--&Y & High     & Yes        & None       & \condcol{Very High} \\
insert sort        & -- &--&Y & Medium   & Yes        & None       & \condcol{High} \\
janne complex      & -- &Y &--& Low      & Yes        & None       & \condcol{Medium} \\
jfdctint           & I  &--&Y & Medium   & Yes        & None       & \condcol{High} \\
lcdnum             & I  &Y &--& Medium   & Yes        & None       & \condcol{High} \\
lms                & FP &Y &--& Medium   & Yes        & None       & \condcol{High} \\
ludcmp             & FP &Y &--& Low      & Yes        & None       & \condcol{Medium} \\
\rowcolor{lightlightgray}
matmult            & I  &--&--& High     & Yes        & None       & \condcol{Very High} \\
minver             & FP &--&--& Low      & Yes        & None       & \condcol{Medium} \\
ndes               & -- &--&Y & Medium   & Yes        & None       & \condcol{High} \\
ns                 & -- &--&Y & Low      & Yes        & None       & \condcol{Medium} \\
nsichneu           & I  &--&Y & Low      & Yes        & None       & \condcol{Medium} \\
prime              & I  &--&--& Low      & Yes        & None       & \condcol{Medium} \\
qsort-exam         & -- &Y &Y & High     & Yes        & None       & \condcol{Very High} \\
qurt               & FP &Y &--& Medium   & Yes        & None       & \condcol{High} \\
recursion          & -- &Y &--& Low      & Yes        & None       & \condcol{Medium} \\
select             & -- &Y &Y & Medium   & Yes        & None       & \condcol{High} \\
sqrt               & FP &Y &--& Medium   & Yes        & None       & \condcol{High} \\
st                 & -- &--&--& Low      & Yes        & None       & \condcol{Medium} \\
statemate          & -- &Y &--& Low      & Possibly   & None       & \condcol{Low} \\
ud                 & I  &Y &--& Low      & Yes        & None       & \condcol{Medium} \\
\midrule
&&&&&&&\\
\bf MediaBench &&&&\\
\midrule
cjpeg              & FP &--&--& Medium   & No         & Major      & \condcol{Low}  \\
djpeg              & FP &--&--& Medium   & No         & Major      & \condcol{Low}  \\
h263dec            & -- &Y &Y & Low      & No         & Major      & \condcol{Very Low}  \\
h263enc            & -- &Y &Y & Low      & No         & Major      & \condcol{Very Low}  \\
h264dec            & -- &Y &Y & Low      & No         & Major      & \condcol{Very Low}  \\
h264enc            & -- &Y &Y & Low      & No         & Major      & \condcol{Very Low}  \\
jpg2000dec         & FP &--&--& Medium   & No         & Major      & \condcol{Low}  \\
jpg2000enc         & FP &--&--& Medium   & No         & Major      & \condcol{Low}  \\
mpeg2dec           & -- &Y &Y & Low      & No         & Major      & \condcol{Very Low}  \\
mpeg2enc           & -- &Y &Y & Low      & No         & Major      & \condcol{Very Low}  \\
mpeg4dec           & -- &Y &Y & Low      & No         & Major      & \condcol{Very Low}  \\
mpeg4enc           & -- &Y &Y & Low      & No         & Major      & \condcol{Very Low}  \\
\midrule
&&&&&&&\\
\bf OpenBench          & --  &--&--& Medium   & No         & Major      & \condcol{Low}     \\
\midrule
&&&&&&&\\
\bf Livermore Loops    & FP &--&--& Low      & Possibly   & None       & \condcol{Low}  \\
\midrule
&&&&&&&\\
\bf LINPACK    & FP &Y&Y& Low      & No   & Major       & \condcol{Very Low}  \\
\midrule
&&&&&&&\\
\bf SPEC2006\footnotemark[2]   & -- &--&--& Low      & No   & Major       & \condcol{Very Low}  \\
\end{supertabular}
\footnotetext[1]{These benchmarks are too small to be useful, their final suitability is adjusted to reflect this.}
\footnotetext[2]{These benchmarks were not available, as this is not a free benchmark suite.}
\end{center}

\end{document}